\documentclass[epjCONF,columns]{svjour} %for 2 columns format
\usepackage{graphics}
\usepackage[varg]{txfonts} % Times fonts
\usepackage[latin1]{inputenc}
\usepackage{subfigure}
\usepackage{slashed}
\newcommand{\MET}{E_T^{miss}}

\usepackage{lineno}

\session-title{Hadron Collider Physics Symposium 2011}
\begin{document}
%\linenumbers
%
\title{Search for 1st-Generation Leptoquarks Using the ATLAS Detector}
\author{John Stupak III\thanks{\email{john.stupak@cern.ch}} on behalf of the ATLAS Collaboration}
\institute{SUNY Stony Brook, Stony Brook, NY}
\abstract{
A search for the pair-production of scalar leptoquarks in $1\textrm{ fb}^{-1}$ of 7 TeV ATLAS data recorded at the LHC is presented. Leptoquarks are hypothetical color-triplet bosons which carry both quark and lepton flavor, and thus decay to a quark and a lepton, unlike any of the Standard Model particles. Leptoquarks arise from many beyond the Standard Model theories. The channels examined in this analysis require at least one leptoquark decay to an electron, which includes the final states eejj and e$\nu$jj. No excess of events is observed, thus limits on allowed leptoquark masses are determined.  We exclude at 95\% confidence level the production of first-generation scalar leptoquarks with mass $m_{LQ}<$ 660 (607) GeV when assuming a branching fraction of leptoquark decay to an electron of 1.0 (0.5).
} %end of abstract
\maketitle
%
%%%%%%%%%%%%%%%%%%%%%%%%%%%%%%%%%%%%%%%%%%%%%%%%%%%%%%%%%%%%%%%%%%%
\section{Introduction}
\label{intro}
Similarities between the generations of leptons	and quarks in the Standard Model (SM) suggest that they might be part of some symmetry at energy scales larger than the electroweak symmetry breaking scale.  Numerous beyond the Standard Model theories include color-triplet gauge bosons known as leptoquarks (LQ), which would mediate the interaction between leptons and quarks~\cite{genLQ}.  Leptoquarks carry both lepton and baryon number, as well as fractional electric charge.  The leptoquark Yukawa coupling $\lambda_{LQ-\ell-q}$,  spin (0 or 1), and branching ratio to charged leptons $\beta$ are model dependent.  Assuming $\lambda_{LQ-\ell-q}$ is of the order of the electroweak coupling strength, leptoquarks are predominantly produced in pairs via gluon fusion and quark-antiquark annihilation.  The cross section for pair-production of scalar leptoquarks depends only on the unknown LQ mass and has been calculated up to next-to-leading order~\cite{kramer}.  To accommodate experimental constraints on flavor changing neutral currents and lepton and baryon number violation,  it is assumed that leptoquarks couple only to quarks and leptons of a single generation.  Thus, they are classified as first-, second-, or third-generation by the generation of leptons to which they couple.  Lower mass bounds for first-generation scalar leptoquarks already exist from searches for leptoquarks produced in pairs at the LHC~\cite{CMS}~\cite{ATLAS}, Tevatron~\cite{DZero}, and LEP~\cite{OPAL}.  Limits on single LQ production come from HERA~\cite{HERA} and other experiments~\cite{PDG}.  

Here we summarize a search for pair-produced, first-generation scalar leptoquarks in pp collisions at $\sqrt{s}$=7TeV, with the ATLAS detector~\cite{thePaper}.  The search uses a dataset corresponding to an integrated luminosity of $1.030\pm0.035\textrm{ fb}^{-1}$.  Two final states are investigated.  One in which both leptoquarks decay to an electron and a quark giving a final state with two electrons and two jets, and another in which only one leptoquark decays to an electron and a quark and the other decays to an electron-neutrino and a quark, giving a final state with one electron, two jets, and missing transverse energy, denoted $\MET$.

\begin{figure*}[!htbp]
	\begin{center}
	\subfigure[]{\resizebox{0.75\columnwidth}{!}{\includegraphics{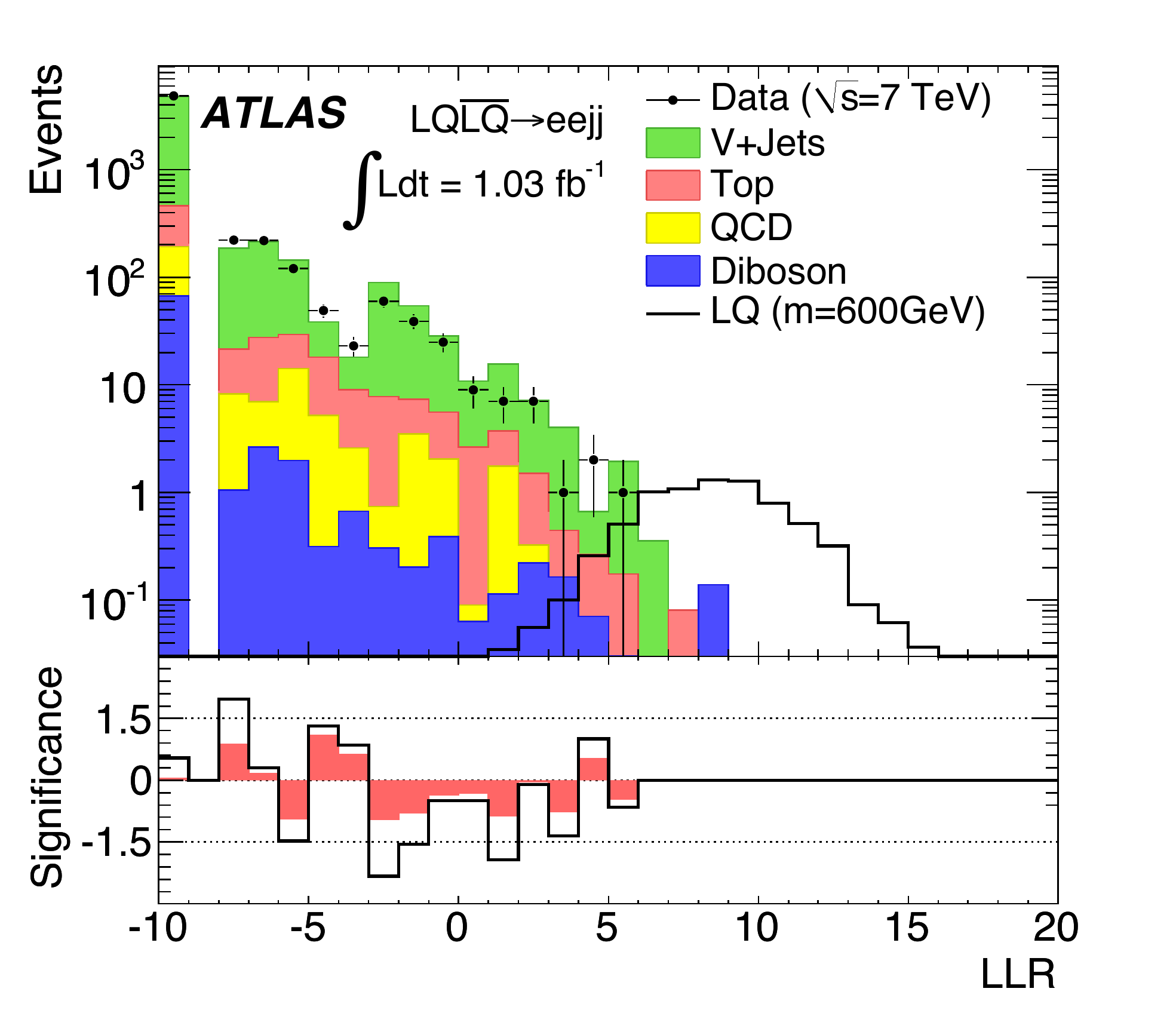}}}
	\subfigure[]{\resizebox{0.75\columnwidth}{!}{\includegraphics{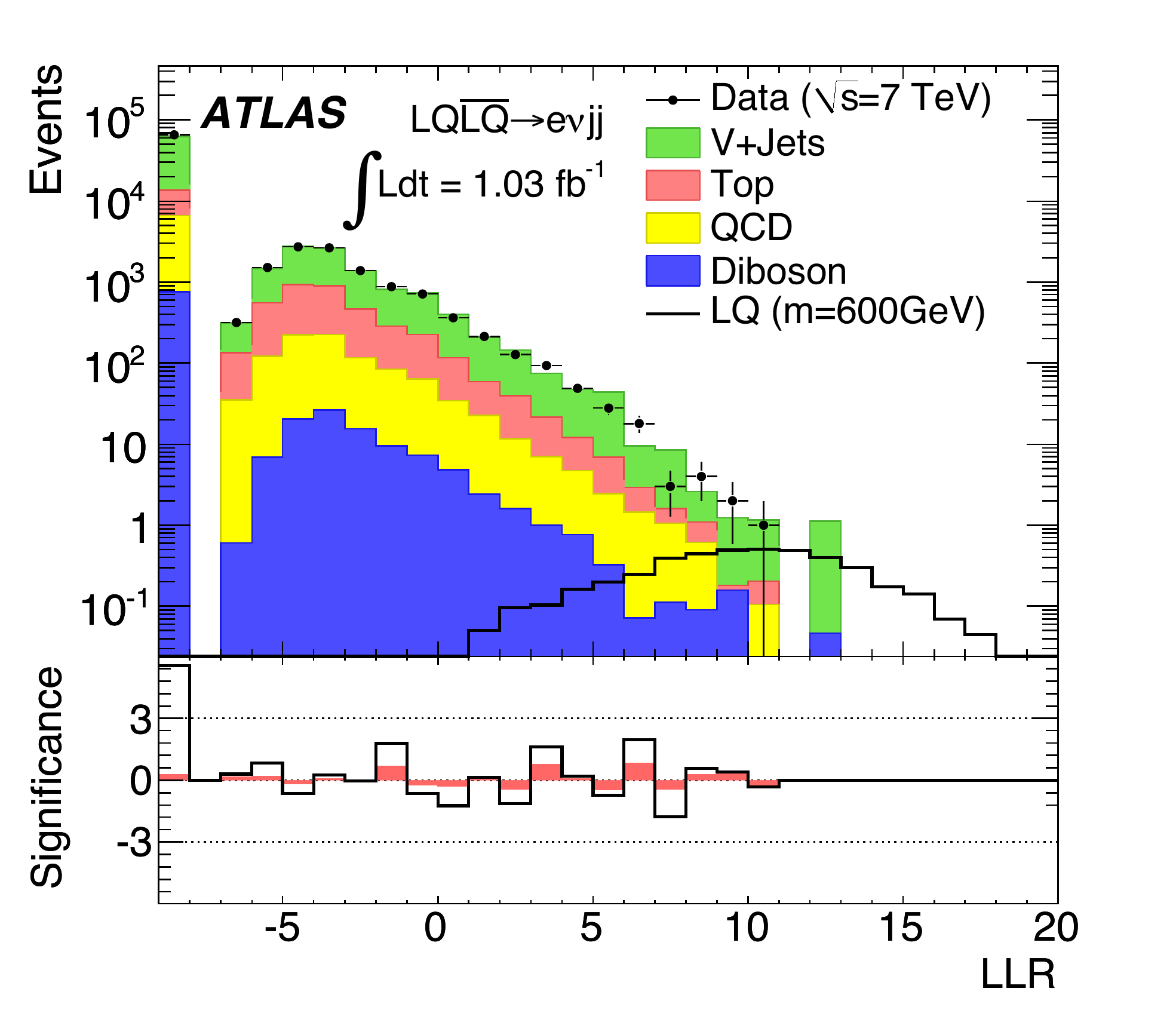}}}
	\end{center}
\caption{LLR distributions (a) for the eejj and (b) for the e$\nu$jj channels. The data are indicated with the points and the filled histograms show the SM background. The QCD background is estimated from data, while the other background contributions are obtained from simulated samples as described in Section~\ref{sec:Background}. The LQ signal corresponding to a LQ mass of 600 GeV is indicated by a solid line, and is normalized assuming $\beta$ = 1.0 (0.5) in the eejj (e$\nu$jj) channel. The lowest bin corresponds to background events in regions of the phase space for which no signal events are expected. The solid line (band) in the lower plot shows the Gaussian statistical (statistical + systematic) significance between data and the prediction.}
\label{fig:LLR}   
\end{figure*}

%%%%%%%%%%%%%%%%%%%%%%%%%%%%%%%%%%%%%%%%%%%%%%%%%%%%%%%%%%%%%%%%%%%%%%%%%
\section{Simulated Samples}
\label{sec:SimSamp}
Monte Carlo samples are used to devise selection criteria, as well as determine background predictions for all backgrounds (excluding the QCD multi-jet background).  These samples are processed through the full GEANT4~\cite{GEANT1} based ATLAS detector simulation~\cite{GEANT2}, followed by the same reconstruction algorithms as used for collision data.  Corrections are made to the simulated samples to ensure a good description of the energy scale and resolution, as well as the trigger reconstruction efficiencies.  The smaller backgrounds are scaled to their cross sections, whereas the normalizations of the major backgrounds are derived from data.  Details of this procedure are discussed in Section~\ref{sec:Background}.

%%%%%%%%%%%%%%%%%%%%%%%%%%%%%%%%%%%%%%%%%%%%%%%%%%%%%%%%%%%%%%%%%%%%%%%%%
\section{Object Identification and Event Selection}
\label{sec:EventSel}

Electrons are required to have a transverse energy $E_T>30$ GeV and fall within a well instrumented region of the detector.  To reduce the contribution from hadrons, electrons are required to have $E_T^{0.2}/E_T<0.1$, where $E_T^{0.2}$ is the transverse energy in a cone of radius $\Delta R=\sqrt{(\Delta\eta^2)+(\Delta\phi^2)}=0.2$ centered on the electron track, excluding the electron contribution.  Jets are reconstructed from energy deposits in the calorimeter using the anti-$k_T$ algorithm~\cite{antiKt} with a distance parameter of 0.4.  Jets are further required to satisfy $E_T>30$ GeV and $|\eta|<2.8$, and be well separated from all electrons $\Delta R>0.4$.  The presence of neutrinos is inferred from the $\MET$ in the event.

For both final states, event selection requirements are defined with high signal efficiency, yet dominated by SM backgrounds.  In both channels, an electron trigger is required, which was $\sim$100\% efficient over the course of data taking.  Also, at least 3 tracks must be associated to the primary vertex and the event must contain at least two jets.  In the eejj channel, exactly two electrons are required and the invariant mass of the pair $m_{ee}$ must be at least 40 GeV.  In the e$\nu$jj channel, exactly one electron is required and a muon veto is applied.  The $\MET$ is required to exceed 30 GeV, and must be well separated from the two leading jets $\Delta\phi(\MET,j_i)>4.5\left(1-\frac{\MET}{45\mathrm{ GeV}}\right)$.  Finally, the transverse mass $m_T(e,\MET)=\sqrt{2\cdot p_T^e\cdot\MET\cdot[1-\cos\Delta\phi(e,\MET)]}$ must be greater than 40 GeV.  These requirements yield a signal acceptance of $\sim$70\% in both channels for an LQ signal assuming $m_{LQ}=600$ GeV.

%%%%%%%%%%%%%%%%%%%%%%%%%%%%%%%%%%%%%%%%%%%%%%%%%%%%%%%%%%%%%%%%%%%%%%%%%
\section{Background Modeling}
\label{sec:Background}

A total of five control regions (CR) are defined with negligible signal contamination and enhanced concentrations of the major backgrounds, either V+Jets (V=W,Z) or $\mathrm{t\overline{t}}$.  They are defined by exploiting differences between signal and background kinematics.  In the eejj channel, the jet multiplicity, electron pair invariant mass $m_{ee}$, and $S_T=p_T^{j_1}+p_T^{j_2}+p_T^{e_1}+p_T^{e_2}$ are used, while in the e$\nu$jj channel the jet multiplicity, transverse mass $m_T(e,\MET)$, and $S_T=p_T^{jet_1}+p_T^{jet_2}+p_T^{electron}+\MET$ are used.  

The QCD contribution is estimated by fully data-driven methods.  A shape template is derived from a QCD enhanced sample.  Depending on the CR, either the $\MET$ or $m_{ee}$ distribution is used to perform a log-likelihood fit.  The relative fraction of QCD compared to all other backgrounds is allowed to float, while the sum is constrained to equal that of the data.  The amount of QCD which minimizes the log-likelihood is taken as the QCD normalization.  This is done separately in each channel, independently for all CRs.  Only the QCD normalization is affected by the outcome of the fitting procedure, not those of the other backgrounds.

The CRs are also used to determine the normalizations of the  V+jets and $\mathrm{t\overline{t}}$ backgrounds.  For each of the major backgrounds, an overall scale factor is determined by minimizing the $\chi^2$ between the predicted and observed yields in each of the CRs.  Correlations amongst the various CRs are taken into account, where necessary.  CR background predictions and data yields are found to agree within systematic uncertainties.

The selected event yield predictied by the SM and observed in data agree within systematic uncertainties as well.  In the eejj (e$\nu$jj) channel we expect $5600\pm1000$ ($74000\pm11000$) events from SM processes and $7.5\pm0.5$ ($4.5\pm0.2$) from a LQ signal assuming $m_{LQ}=600$ GeV, and we observe 5615 (76855) events in data.

%%%%%%%%%%%%%%%%%%%%%%%%%%%%%%%%%%%%%%%%%%%%%%%%%%%%%%%%%%%%%%%%%%%%%%%%%
\section{Log-Likelihood Ratio Discriminant}
\label{sec:LLR}

A log-likelihood ratio (LLR) method is used to separate signal and background.  A set of discriminating variables are chosen which differ considerably for signal and background.  In the eejj channel, $m_{ee}$, $S_T$, and the average LQ mass resulting from the best (electron,jet) combination in each event are used.  In the e$\nu$jj channel, $m_T(e,\MET)$, $S_T$, and the invariant and transverse LQ masses resulting from the best (electron,jet) and ($\MET$,jet) combination in each event are used.  The best combination is the pairing which minimizes the mass difference between the two leptoquarks.

Probability distribution functions (PDF) are formed for each of these discriminating variables, separately for signal and background.  From these PDFs, joint likelihoods $L_S=\displaystyle\prod\limits_{i} P_S^i\left(x_j\right)$ and $L_B=\displaystyle\prod\limits_{i} P_B^i\left(x_j\right)$ are formed, where $P_s^i(x_j)$ and $P_B^i(x_j)$ are the probabilities of the i-th input variables having a value of $x_j$ in signal and background events, respectively.  $L_S$ distributions are formed for each LQ mass hypothesis tested, allowing a mass-dependent optimization.  From these joint likelihoods, the log-likelihood ratio $LLR=\log\left(L_S/L_B\right)$ is formed, which is shown in Figure~\ref{fig:LLR} for both channels.

\begin{table}[htbp]
\caption{The predicted and observed yields in a signal enhanced region defined by requiring $LLR>0$ for both channels.  Background predictions are scaled as described in Section~\ref{sec:Background}. The eejj (e$\nu$jj) channel signal yields are computed assuming $\beta=1.0\,\,(0.5)$. Statistical and systematic uncertainties added in quadrature are shown.}
\label{tab:EvSelYield}
\begin{tabular}{lccccc}
\hline\noalign{\smallskip}
Source  & \multicolumn{2}{c}{eejj Channel} & \multicolumn{2}{c}{e$\nu$jj Channel}\\
			& 400 GeV & 600 GeV & 400 GeV & 600 GeV \\ 
\noalign{\smallskip}\hline\noalign{\smallskip}
$W$+jets                   & ---              & ---              & $1500 \pm 670$ & $ 670 \pm 210$ \\
$Z$+jets & $ 98 \pm 53$ & $ 26 \pm 14$ & $  45 \pm 41$ & $  18 \pm 19$ \\
$\mathrm{t\overline{t}}$ & $ 15 \pm  9$ & $  4.6 \pm  2.2$ & $ 430 \pm 180$ & $ 150 \pm  38$ \\
Single $t$ & $  1.4 \pm  0.9$ & $  0.7 \pm  0.4$ & $  53 \pm  19 $ & $  23 \pm 4$ \\
Dibosons& $  1.5 \pm  0.8$ & $  0.7 \pm  0.3$ & $  25 \pm  11$ & $  11 \pm 2$ \\
QCD& $  9.2 \pm  4.5$ & $  2.3 \pm  1.5$ & $ 170 \pm  35$ & $  75 \pm  15$ \\
\noalign{\smallskip}\hline\noalign{\smallskip}
Total& $120 \pm 55$ & $ 34 \pm 14$ & $2200 \pm 690$ & $ 950 \pm 220$ \\
Data & $ 82$            & $  22$           & $ 2207$           & $   900 $          \\
\noalign{\smallskip}\hline\noalign{\smallskip}
LQ & $120 \pm 8$ & $ 7.5 \pm 0.5$ & $  69 \pm 4$  & $   4.5 \pm  0.2$  \\
\noalign{\smallskip}\hline
\end{tabular}
\end{table}

%%%%%%%%%%%%%%%%%%%%%%%%%%%%%%%%%%%%%%%%%%%%%%%%%%%%%%%%%%%%%%%%%%%%%%%%%
\section{Systematics}
\label{sec:sys}

Systematic uncertainties affect both the background normalizations and the shapes of the distributions which serve as inputs to the LLR. We consider systematic uncertainties from a variety of sources.  Energy scale and resolution uncertainties are evaluated for electrons and jets.  Electron trigger, reconstruction, and identification efficiency uncertainties are included as well.  The V+jets and $\mathrm{t\overline{t}}$ production model uncertainties are evaluated using alternative MC generators and by varying generator parameters.  The systematic uncertainty on the QCD background is estimated by using alternative distributions for the fitting procedure described in Section~\ref{sec:Background}.  A systematic uncertainty is applied to those samples whose normalization is not taken from data, in order to account for integrated luminosity uncertainty.  Systematic uncertainties are also included to account for the modeling of initial and final state radiation, and the choices of hadronization/factorization scale and parton distribution function in signal MC.

%%%%%%%%%%%%%%%%%%%%%%%%%%%%%%%%%%%%%%%%%%%%%%%%%%%%%%%%%%%%%%%%%%%%%%%%%
\section{Results}
\label{sec:Results}

\begin{figure}[htbp]
	\begin{center}
	\resizebox{0.75\columnwidth}{!}{\includegraphics{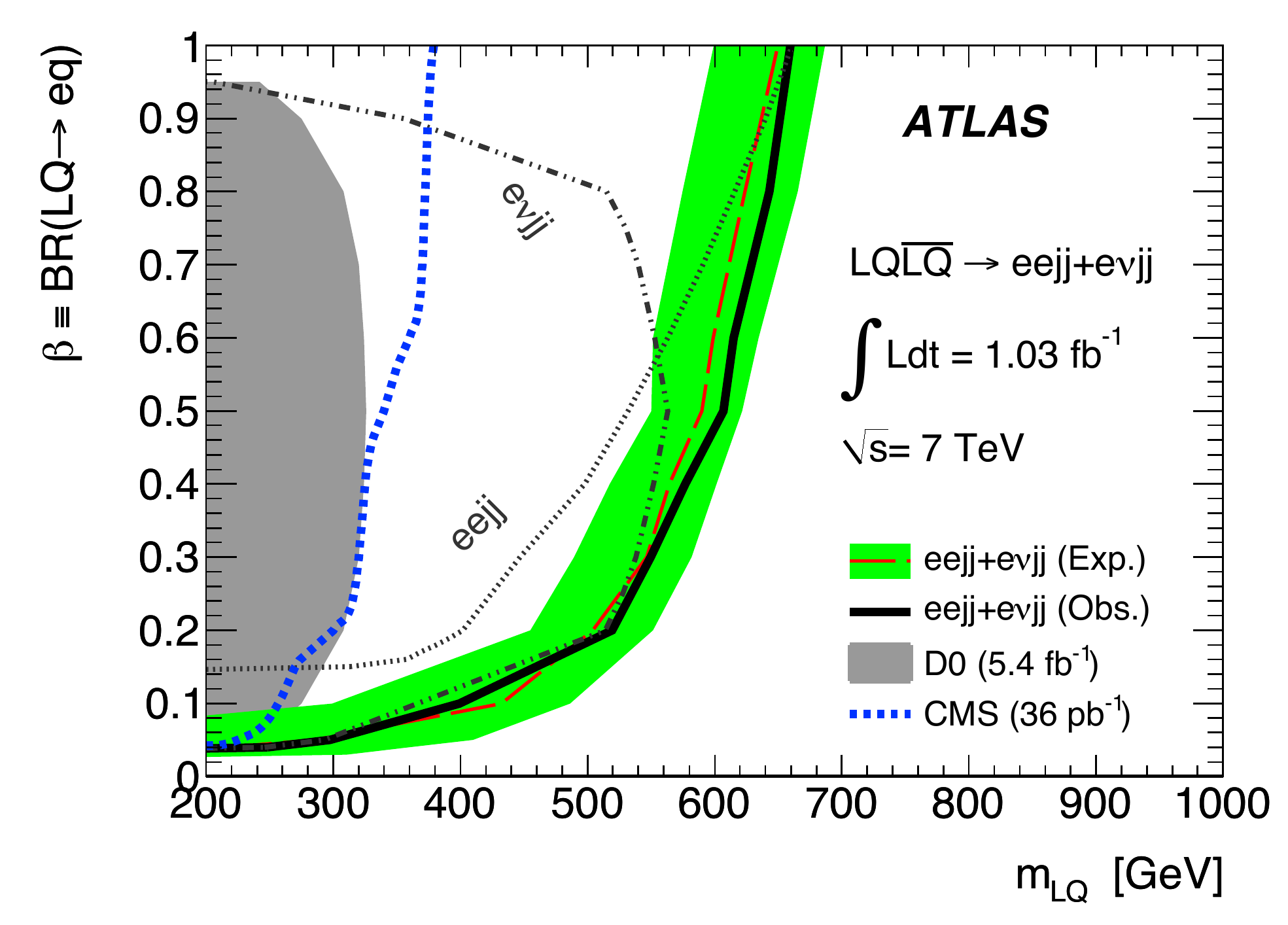}}
	\caption{95\% CL exclusion region resulting from the combination of the two channels shown in the $\beta$ versus leptoquark mass plane. The shaded area indicates the D0 exclusion limit, while the thick dotted line indicates the CMS exclusion. The dotted and dotted-dashed lines indicate the individual limits for the eejj and the e$\nu$jj channels, respectively. The combined expected limit is indicated by the dashed line, together with the systematics band resulting from the $\pm1\sigma$ variation. The combined observed limit is indicated by the solid black line.}
	\label{fig:comb_ex}
	\end{center}
\end{figure}

In neither channel do we observe an excess of events at high LLR where the signal is expected, indicating no evidence of scalar LQ pair-production.  The observed and predicted event yields requiring $LLR>$ 0 are shown in Table~\ref{tab:EvSelYield}. In the absence of a signal, 95\% CL upper limits on the LQ pair-production cross section are set using a modified frequentist $CL_S$ method  based on a Poisson log-likelihood ratio statistical test~\cite{CLS1}~\cite{CLS2}. Statistical and systematic uncertainties are treated as nuisance parameters with a Gaussian probability density function, and the full LLR distribution is considered.  The cross section upper bounds are reinterpreted in the $\beta$ versus $m_{LQ}$ plane, yielding the exclusion limits shown in Figure~\ref{fig:comb_ex}.

%%%%%%%%%%%%%%%%%%%%%%%%%%%%%%%%%%%%%%%%%%%%%%%%%%%%%%%%%%%%%%%%%%%%%%%%%
\section{Conclusions}
\label{sec:Conclusions}

We report on a search for pair-production of first-generation scalar leptoquarks at ATLAS using a data sample corresponding to an integrated luminosity of 1.03 
$\textrm{fb}^{-1}$. 
No excess over SM background expectations is observed in the data in the signal enhanced region, and 95\% CL upper bounds on the production cross section are determined. These are translated into lower observed (expected) limits on leptoquark masses of 
$m_{LQ}>$ 660 (650) GeV and $m_{LQ}>$ 607 (587) GeV when assuming its branching fraction to an electron is equal to 1.0 and 0.5, respectively. These are the most stringent limits to date arising from direct searches for leptoquarks.

%\begin{figure}
%% Use the relevant command for your figure-insertion program
%% to insert the figure file.
%% For example, with the option graphics use
%%\resizebox{0.75\columnwidth}{!}{%
%%  \includegraphics{fig1.eps} }
%\caption{Please write your figure caption here.}
%\end{figure}
%\begin{table}
%\caption{Please write your table caption here.}
%\begin{tabular}{lll}
%\hline\noalign{\smallskip}
%\end{tabular}
%\end{table}
%

\end{document}